\documentclass[aps,pra,showpacs,%
	amsmath,%
	reprint%
	]{revtex4-1}

\newcommand{\set}[1]{\left\{ #1 \right\}}
\newcommand{\pdfrac}[2]{\frac{\partial #1}{\partial #2}}
\def\toexp{\mathop{\rm exp}}

\newcommand{\AntiTexp}{\toexp_{\rightarrow}}

\begin{document}
\preprint{{\tt AT 2010-07-10-1}}

\title{Quantum anholonomies in time-dependent Aharonov-Bohm rings}

\author{Atushi Tanaka}
\homepage[]{\tt http://researchmap.jp/tanaka-atushi/}
\affiliation{Department of Physics, Tokyo Metropolitan University,
  Hachioji, Tokyo 192-0397, Japan}

\author{Taksu Cheon}
\homepage[]{\tt http://researchmap.jp/T_Zen/}
\affiliation{Laboratory of Physics, Kochi University of Technology,
  Tosa Yamada, Kochi 782-8502, Japan}

\date{\today}

\begin{abstract}
Anholonomies in eigenstates are studied through time-dependent
variations of a magnetic flux in an Aharonov-Bohm ring.  The
anholonomies in the eigenenergy and the expectation values of
eigenstates are shown to persist beyond the adiabatic regime.  The
choice of the gauge of the magnetic flux is shown to be crucial to
clarify the relationship of these anholonomies to the eigenspace
anholonomy, which is described by a non-Abelian connection in the
adiabatic limit.
\end{abstract}


\maketitle

\section{Introduction}
The parametric dependence of eigenenergies and eigenfunctions
of a Hamiltonian 
offers a key to understanding hierarchical quantum systems, e.g.,
the band theory for solids and the Born-Oppenheimer 
approximation for molecules.
Recently, it was shown that there are examples where 
eigenenergies and eigenfunctions are multiple-valued functions 
of a parameter, and cyclic adiabatic variations of the parameter transform 
one eigenstate into another~\cite{Cheon-PLA-248-285}.
Such phenomena are called eigenvalue and eigenspace anholonomies, 
or exotic quantum holonomy~\cite{Cheon-EPL-85-20001,TANAKA-AP-85-1340}.
Its applications to adiabatic manipulation of quantum states, including 
adiabatic quantum computation~\cite{Farhi-quant-ph-0001106}, were also
examined recently~\cite{Tanaka-PRL-98-160407,Tanaka-PRA-81-022320}.
The eigenspace anholonomy was also examined through 
the adiabatic Floquet theory~\cite{Viennot-JPA-42-395302}.
Up to now, exotic quantum holonomies were considered only in their strictly 
adiabatic limit. 
It is important to establish the robust existence of exotic quantum 
holonomy in the non-adiabatic realm, particularly in light of its potential
experimental realization and also its potential use in quantum computation.

In this article, we present a first attempt to extend the
concept of exotic quantum holonomy into time-dependent parametric
motion away from the adiabatic limit. 
Throughout this work, we examine a charged particle in a 
one-dimensional ring, where a magnetic flux is applied.
Although this system has been extensively investigated, as it
exhibits the Aharonov-Bohm effect~\cite{Aharonov-PR-115-485} and
the ``persistent current''~\cite{Buettiker-PLA-96-365},
there has been no serious argument on exotic holonomies, as far as 
the authors are aware.
It turns out that this system, with its simplicity, allows us to
establish the existence of eigenvalue and eigenspace anholonomies in
a time-dependent system beyond the adiabatic variation of parameters.

At the same time, we find that the freedom of the choice of
the gauge of the magnetic flux in the Aharonov-Bohm ring offers 
a subtle problem on the eigenspace anholonomy.
In all the conventional examples, eigenvalue and eigenspace anholonomies 
appear in consort, which, at first sight, seems
rather natural due to the correspondence of eigenvalues and
eigenfunctions in Hermite operators.
However, since eigenfunction itself is not an observable, there is
no reason that it should follow the eigenvalue in its anholonomic variation,
although any observable quantities calculated from it obviously 
has to show the anholonomy.

In fact, in the system we consider,
the gauge transformation of the magnetic field critically changes
the way the eigenstates vary under cyclic variation of the relevant parameter.
As a result, the variation of the eigenfunction
may not display the expected anholonomy
accompanying the eigenvalue anholonomy.
Indeed, it is shown that the eigenspace 
anholonomy appears only under a suitable choice of the vector
potential of the magnetic field. 

This paper is organized as follows. In Sec.~\ref{sec:singleValued},
we introduce the Aharonov-Bohm ring, and examine its properties under
the gauge where the wavefunction is periodic in the ring. 
Due to the simplicity of this model, it 
suffices to examine how the eigenvalues and eigenfunctions depend on 
the magnetic flux, even when the magnetic flux is time-dependent. 
An example where the anholonomy in eigenenergies
does not accompany any anholonomy in eigenfunctions is shown.
In Sec.~\ref{sec:BYgauge}, we examine the time evolution under 
the gauge devised by Byers and Yang where the gauge potential 
of the magnetic flux is removed from the Hamiltonian~\cite{Byers-PRL-7-46}.
The eigenspace and the eigenenergy anholonomies are shown to be synchronized.
Both adiabatic and non-adiabatic variations of the magnetic flux are examined.
In Sec.~\ref{sec:nonAbelian}, the parametric dependence of eigenfunctions
under the Byers-Yang gauge is examined by a non-Abelian gauge 
connection, which is the crucial element that controls
quantum anholonomies~\cite{Cheon-EPL-85-20001}.
In Sec.~\ref{sec:adiabaticBYgauge}, we examine the adiabatic time evolution in
the Byers-Yang gauge to clarify the eigenspace anholonomy
in terms of a non-Abelian gauge potential, following a recent formulation
for the exotic holonomy~\cite{Cheon-EPL-85-20001}. 
It is also shown that the dynamical phase in the Byers-Yang gauge 
has a geometric part.

\section{Aharonov-Bohm ring}
\label{sec:singleValued}
We consider a quantum particle on a ring threaded by a magnetic flux 
of variable strength.
First of all, we explain the choice of the vector potential where 
wavefunctions satisfy the periodic boundary condition in the ring.
The particle, whose charge is $q$,
is described by the Hamiltonian
\begin{equation}
  \label{eq:HamiltonianInPeriodicGauge}
  H
  \equiv\frac{1}{2}\left[\frac{1}{i}\pdfrac{}{x} -q A(x)\right]^2
  ,
\end{equation}
where $x$ denotes the position of the particle on the ring 
($0\le x < L$), and $A(x)$ is the tangent component of the 
vector potential at the ring.
We choose the units that $\hbar$ and the mass of the particle are
unity.
We assume that the deforming effect of ring curvature can be neglected.
Also, we assume that the ring is so clean that the scalar potential can be
ignored.
Let us denote $\Phi$ be the magnetic flux, normalized by 
the flux quantum $2\pi/q$, applied through the ring, i.e.,
\begin{equation}
  \Phi \equiv \frac{1}{2\pi/q}\int_0^L A(x)dx
  .
\end{equation}
For the sake of simplicity, we choose $A(x) = 2\pi\Phi/qL$.

Once the gauge of the electromagnetic field is suitably chosen,
this system is periodic for the normalized magnetic flux $\Phi$ 
with a period $1$~\cite{Byers-PRL-7-46}, as will be shown in 
the next section. Accordingly, the path $C$ where $\Phi$ is increased 
by its unit, say $\Phi'$ to $\Phi'+1$, may be regarded as a closed one. 
It will also be shown that the anholonomies in the eigenvalues and 
the eigenspaces occur for the cycle $C$ in the next section.

However, whether $C$ is closed or open essentially depends on 
the choice of the gauge. In particular, $C$ must be regarded as 
open under the present choice of the gauge, because the Hamiltonian 
[Eq.~\eqref{eq:HamiltonianInPeriodicGauge}] is not periodic for $\Phi$.
In the following, we show that gauge invariant quantities, such as 
eigenenergies and expectation values of eigenstates, depend
on $\Phi$ in an aperiodic manner. This reflects the anholonomies
that appear in the gauge where $C$ is periodic.

We examine the time evolution of this system during the 
increment of the magnetic flux $\Phi$ by its period, i.e., 
from $\Phi'$ at $t=t'$ to $\Phi''=\Phi'+1$ at $t=t''$ ($>t'$).

When the system is initially in an eigenstate,
the system stays in the eigenstate regardless of the
speed of the variation of the magnetic field.
This is because  eigenfunctions of 
$H[\Phi(t)]$ can be chosen to be time-independent at any instant.
Indeed, a normalized eigenfunction of $H$ for a given $\Phi$ is
\begin{equation}
  \label{eq:eigenfunctionInPeriodicGauge}
  \psi_k (x) = \frac{1}{\sqrt{L}}e^{i2\pi k x / L}
  ,
\end{equation}
for an integer $k$.
On the other hand, the $k$-th eigenenergy for a given $\Phi$ is
\begin{equation}
  \label{eq:Ek}
  E_k(\Phi) = \frac{1}{2}\left[\frac{2\pi(k -\Phi)}{L}\right]^2
  .
\end{equation}
Now it is straightforward to obtain the solution of the
time-dependent Schr\"odinger equation 
\begin{equation}
  i\pdfrac{}{t}\Psi(t, x) = H[\Phi(t)]\Psi(t, x)
\end{equation}
with the initial condition $\Psi(t', x) = \psi_k(x)$.
At the end of the path $C$, we obtain 
\begin{equation}
  \label{eq:timeEvolutionInThePeriodicGauge}
  \Psi(t'', x) = e^{i\gamma_{\rm D}}\psi_k(x)
  ,
\end{equation}
where 
\begin{equation}
  \label{eq:DefDynamicalPhase}
  \gamma_{\rm D}\equiv-\int_{t'}^{t''} E_k[\Phi(t)] dt 
\end{equation}
is the dynamical phase~\cite{Berry-PRSLA-392-45}.
Hence, it is sufficient examine the $\Phi$-dependence of 
eigenvalues and eigenstates to elucidate anholonomies 
both for adiabatic and nonadiabatic variations of $\Phi$
along $C$.

We examine how eigenenergies changes along $C$.
From Eq.~\eqref{eq:Ek},
the energy spectrum 
$
\sigma[H(\Phi)] \equiv \set{E_k(\Phi)}_{k=-\infty}^{\infty}
$
is periodic in $\Phi$ with a period $1$,
i.e.,
$
  \sigma[H(\Phi+1)] = \sigma[H(\Phi)]
  .
$
In this sense, $C$ is regarded to be closed for $\sigma[H(\Phi)]$.
On the other hand,
each eigenvalue does not obey the periodicity, e.g.,
\begin{equation}
  \label{eq:eigenenergyAnholonomy}
  E_k(\Phi+1) = E_{k-1}(\Phi)
  ,
\end{equation}
which indicates the presence the eigenenergy anholonomy 
for the
path $C$.

The eigenenergy anholonomy implies that 
an adiabatic increment of $\Phi$ along $C$
transports the $k$-th eigenstate to the $k-1$-th eigenstate,
as long as the spectrum degeneracies are not broken due to
perturbations~\cite{Buettiker-PLA-96-365}. 
However, such an argument seems to be inconsistent with the fact 
that the $k$-th 
eigenfunction $\psi_k (x)$ [Eq.~\eqref{eq:eigenfunctionInPeriodicGauge}]
is independent of $\Phi$. In fact, this does not immediately lead
to any contradiction, 
because not only $\psi_k (x)$ but also the ray of $\psi_k (x)$
depend on the gauge of the electromagnetic field,
and thus are not observables. 

To characterize such an anholonomy in an eigenstate, we need to 
focus on the gauge invariant properties of the eigenstate.
The parametric dependence of the expectation values
of observables, which are gauge invariant, are consistent with 
the eigenenergy anholonomy [Eq.~\eqref{eq:eigenenergyAnholonomy}].
For example, the expectation value of the velocity operator 
\begin{equation}
  \label{eq:defv}
  v \equiv \frac{1}{i}\pdfrac{}{x} -q A(x)
  ,
\end{equation}
for the $k$-th eigenfunction $\psi_k (x)$
is
\begin{equation}
  v_k(\Phi) = \frac{2\pi(k-\Phi)}{L}
  .
\end{equation}
Another example is the probability current density
\begin{equation}
  j_k(\Phi) = \frac{2\pi(k-\Phi)}{mL^2}
\end{equation}
for the $k$-th eigenstate. 
They are 
consistent with 
the $\Phi$-dependence of the eigenenergy 
$E_k(\Phi)$ [Eq.~\eqref{eq:eigenenergyAnholonomy}], i.e., 
\begin{equation}
  v_k(\Phi+1) = v_{k-1}(\Phi),
  \quad\text{and}\quad
  j_k(\Phi+1) = j_{k-1}(\Phi)
  .
\end{equation}
Thus it is clear that we cannot extract
any $\Phi$-dependence of the $k$-th eigenstate from $\psi_k (x)$
if we examine only the eigenfunction itself, and overlook
the expectation values of observables.
Namely, the eigenspace anholonomies of this system
are described by the anholonomies in all collections of 
expectation values.

A remark on the gauge dependence of the Hermite operators is in order.
The position operator $x$ is gauge invariant, though its
expectation value is useless in examining the anholonomy of
the present system, as the expectation value for an eigenstate
happens to be independent of $\Phi$. Although the momentum
operator $-i\partial_x$ is also gauge invariant, its expectation
value is gauge dependent. The velocity operator (or covariant
momentum operator) $v$ [Eq.~\eqref{eq:defv}] is gauge covariant 
and offers a key to identify the anholonomy, as shown above.

This result suggests that the standard treatment
of the eigenspace anholonomy~\cite{Cheon-EPL-85-20001} 
is inapplicable to the present case,
since the prescription in Ref.~\cite{Cheon-EPL-85-20001}
essentially depends on the choice of the vector potential 
of the electromagnetic field.
Is there any way to reconcile the present argument and
the standard treatment
of the eigenspace anholonomy~\cite{Cheon-EPL-85-20001}?

In the following, we shall show that an appropriate choice of
the gauge of the electromagnetic field allows us to employ 
the prescription in Ref.~\cite{Cheon-EPL-85-20001}.
A key is the use of a gauge devised by 
Byers and Yang~\cite{Byers-PRL-7-46},
which provides a tool to elucidate the thermodynamic
properties of Aharonov-Bohm systems.
Under the Byers-Yang gauge, the parametric dependence of
the eigenenergies and the eigenfunctions are associated
with the eigenvalue and the eigenspace anholonomies.

\section{Quasi-periodic gauge}
\label{sec:BYgauge}
A proper choice of gauge transformation for the magnetic 
flux offers 
a key to resolve the question above. 
We show that, under a suitable gauge,
the eigenvalue and 
eigenspace anholonomies are synchronized.
Furthermore, we show that the Aharonov-Bohm ring offers an example 
that exhibits eigenspace anholonomy in both the adiabatic and 
nonadiabatic regimes.

A gauge transformation devised by Byers and Yang
is defined 
for a wave function ${\psi}(x)$
that satisfies the periodic boundary condition~\cite{Byers-PRL-7-46}:
\begin{align}
  \label{eq:BYgaugeTransformation}
  \tilde{\psi}(x)
  = \exp\left[-iq\int_0^{x}A(x')dx'\right]{\psi}(x)
  .
\end{align}
The resultant wavefunction $\tilde{\psi}(x)$ obeys
a quasi-periodic boundary condition
\begin{equation}
  \label{eq:BynersYangPBC}
  \tilde{\psi}(L)
  = e^{-i2\pi\Phi}\tilde{\psi}(0)
  ,
\end{equation}
which is a manifestation of 
the Aharonov-Bohm effect~\cite{Aharonov-PR-115-485}.
The Hamiltonian for $\tilde\psi (x)$ is
\begin{equation}
  \label{ed:defHstatic}
 \tilde{H}^{\rm static}
 \equiv -\frac{1}{2}\pdfrac{^2}{x^2}
 ,
\end{equation}
where we put the superscript {\it static} to stress that the magnetic flux
is time-independent. As for the time-dependent case, we refer to
Eq.~\eqref{eq:ExactHamiltonianInBYgauge} in Sec.~\ref{sec:adiabaticBYgauge}.

Here we show that this system is periodic in $\Phi$ with period $1$ under 
the Byers-Yang gauge~\cite{Byers-PRL-7-46}.
First, $\tilde{H}^{\rm static}$ [Eq.~\eqref{ed:defHstatic}]
is independent of $\Phi$.
Second, the quasi-periodic boundary condition [Eq.~\eqref{eq:BynersYangPBC}]
itself
is periodic in $\Phi$ with period 1.
Hence the periodicity of the system is evident.
The path $C$ where $\Phi$ is increased by its unit is closed under 
the Byers-Yang gauge.

The most crucial difference between the Byers-Yang gauge
and the previous choice of the magnetic gauge (we call it the periodic gauge
in the following) is 
seen in
eigenfunctions.
A $k$-th eigenfunction of $\tilde{H}^{\rm static}$ in the Byers-Yang gauge is
\begin{equation}
  \label{eq:defTildePsik}
  \tilde{\psi}_k(x; \Phi)
  \equiv
  \frac{1}{\sqrt{L}} 
  \exp\left[i\frac{2\pi (k-\Phi) x}{L} + i\pi\Phi\right]
  ,
\end{equation}
where we choose the second term in the exponent so as to
satisfy the parallel transport condition~\cite{Stone-PRSLA-351-141}
for $\Phi$, i.e., 
\begin{equation}
  \langle \tilde{\psi}_k(x; \Phi), \partial_{\Phi}\tilde{\psi}_k(x; \Phi)
  \rangle = 0.
\end{equation}

We show the eigenspace anholonomy for an adiabatic closed path $C$.
It is sufficient to examine the parametric dependence on the eigenfunction
thanks to the assumption of adiabaticity.
From Eq.~\eqref{eq:defTildePsik}, we have
\begin{equation}
  \label{eq:parametricDependenceOnTildePsi}
  \tilde{\psi}_k(x; \Phi'+1)
  = e^{i\pi}\tilde{\psi}_{k-1}(x; \Phi')
  .
\end{equation}
This means that the adiabatic time evolution of the state vector
whose initial
condition is obtained as $\tilde{\psi}_{k-1}(x; \Phi')$, apart from 
the phase factor.
This immediately indicates the presence of the eigenspace anholonomy.
Namely, the state vector that initially belongs to the $k$-th eigenspace 
is adiabatically transported to the $k-1$-th eigenspace along a periodic 
increment of $\Phi$ by unity.

We extend our analysis beyond the adiabatic regime.
Namely, we examine the time evolution of the wave function 
in the Byers-Yang gauge along 
the time-development of the magnetic flux $\Phi(t)$
from $\Phi'$ at $t = t'$ to $\Phi''$ at $t=t''$.
We here assume that $\Phi(t)$ gently starts and stops at 
$t= t'$ and $t''$, respectively, to ensure 
the applicability of Eq.~\eqref{ed:defHstatic} at both ends.
The initial wavefunction is assumed to be 
the $k$-th eigenfunction of
$\tilde{H}^{\rm static}$ at $t=t'$, i.e.,
\begin{equation}
  \tilde{\psi}_k(x; \Phi')
\end{equation}
In the periodic gauge, the initial wavefunction is
$e^{i\pi\Phi'}\psi_k(x)$, which is an eigenfunction 
of Eq.~\eqref{eq:HamiltonianInPeriodicGauge} for an arbitrary strength 
of the magnetic flux.
Hence, the wavefunction in the periodic gauge at $t=t''$ is 
$e^{i\gamma_{\rm D}}e^{i\pi\Phi'}\psi_k(x)$, where
$\gamma_{\rm D}$ is the dynamical phase, as previously shown in
Eq.~\eqref{eq:DefDynamicalPhase}.
The final wavefunction in the Byers-Yang gauge is
\begin{equation}
  \label{eq:exactTimeEvolutionViaPeriodicGauge}
  e^{i\gamma_{\rm D}}e^{-i\pi(\Phi''-\Phi')}\tilde{\psi}_k(x; \Phi'')
  ,
\end{equation}
which is an eigenstate of $\tilde{H}^{\rm static}$ at $t=t''$.
Note that the final wavefunction, except its dynamical phase 
$\gamma_{\rm D}$, is independent of the precise time 
dependence of $\Phi$.

Here we make a remark on the second 
factor in Eq.~\eqref{eq:exactTimeEvolutionViaPeriodicGauge} 
in light of the adiabatic change of $\Phi$.
For parameters other than the electromagnetic field, the parallel 
transport condition ensures the correspondence of the parametric dependence
of eigenstates and the adiabatic time evolution~\cite{Simon-PRL-51-2167}. 
In our case, although $\tilde{\psi}_k(x; \Phi)$ 
[Eq.~\eqref{eq:defTildePsik}] satisfies the parallel transport condition,
Eq.~\eqref{eq:exactTimeEvolutionViaPeriodicGauge} tells us that
the wavefunction acquires an extra phase $-\pi(\Phi''-\Phi')$.
We shall clarify the origin of the extra phase 
in Sec.~\ref{sec:adiabaticBYgauge},
where we need to carefully 
examine the adiabatic time evolution in the Byers-Yang gauge.

We apply the above result to examine the eigenspace anholonomy 
for the closed path $C$.
From Eq.~\eqref{eq:exactTimeEvolutionViaPeriodicGauge},
we obtain the wavefunction at $t=t''$ as 
\begin{equation}
  \tilde{\psi}^{(C)}_k(x; \Phi')
  \equiv 
  e^{-i\pi}\tilde{\psi}_k(x; \Phi'+1)
  ,
\end{equation}
where the dynamical phase factor is excluded.

In order to compare the final wavefunction $\tilde{\psi}^{(C)}_k(x; \Phi')$
with the initial eigenfunctions,
we look at the parametric dependence of 
$\tilde{\psi}_k(x; \Phi'+1)$ with $\Phi'$.
From Eq.~\eqref{eq:parametricDependenceOnTildePsi}, we obtain
the finial wavefunction
in terms of the initial eigenfunctions as
\begin{equation}
  \label{eq:tildePsiParallelTransported}
  \tilde{\psi}^{(C)}_k(x; \Phi')
  =
  \tilde{\psi}_{k-1}(x; \Phi')
  .
\end{equation}

We are now ready to characterize the eigenspace anholonomy
by the holonomy matrix $M(C)$ whose
$(k'',k')$-th element is the overlapping integral
between $\tilde{\psi}_{k''}(x; \Phi')$ 
and 
$\tilde{\psi}^{(C)}_{k'}(x; \Phi')$~\cite{Cheon-EPL-85-20001,Kim-PLA-374-1958},
i.e.,
\begin{equation}
  \label{eq:holonomyMatrixByParallelTransport}
  M_{k'', k'} (C)
  = \langle\tilde{\psi}_{k''}(x; \Phi'), \;
  \tilde{\psi}^{(C)}_{k'}(x; \Phi')\rangle
  .
\end{equation}
From Eq.~\eqref{eq:tildePsiParallelTransported}, we obtain
\begin{equation}
  \label{eq:holonomyMatrix}
  M_{k'', k'}(C) = \delta_{k'', k'-1}
  .
\end{equation}
Hence $M(C)$ is a permutation matrix, which precisely
describes the eigenspace anholonomy.

We show that all the anholonomies in the expectation values of observable 
can be represented by the eigenspace anholonomy.
The link between the eigenspace anholonomy, and, the anholonomies
in the eigenenergy and the expectation values of eigenstates are 
restored 
in the Byers-Yang gauge.
We emphasize that this result is valid for arbitrary time dependence
of $\Phi(t)$, i.e., both for adiabatic and nonadiabatic ones,
as long as $\Phi(t)$ starts and stops gently enough at 
both ends.

\section{A non-Abelian connection}
\label{sec:nonAbelian}

The argument in the previous section heavily depends on
the choice of the phase factor of $\tilde{\psi}_k(x;\Phi)$
in Eq.~\eqref{eq:defTildePsik}.
We take into account such an arbitrariness by using 
a non-Abelian gauge connection
\begin{equation}
  \label{eq:gaugeConnectionInBYgauge}
  \tilde{A}_{k'',k'}(\Phi)
  \equiv 
  \langle \tilde{\psi}_{k''}(x; \Phi), 
  i\partial_{\Phi}\tilde{\psi}_{k'}(x; \Phi)
  \rangle
  ,
\end{equation}
which is induced by $\tilde\psi_k(x;\Phi)$~\cite{Cheon-EPL-85-20001}.
Note that the term ``gauge'' for Eq.~\eqref{eq:gaugeConnectionInBYgauge}
is analogous, but different from the gauge for the magnetic flux
of the present system.

The non-Abelian gauge connection $\tilde{A}(\Phi)$
describes the infinitesimal
change of basis vectors 
$\{\tilde{\psi}_{k}(x; \Phi)\}_k$~\cite{Filipp-PRA-68-02112}. Namely,
$\tilde{\psi}_{k}(x; \Phi)$ satisfies 
a differential equation
\begin{equation}
  i\pdfrac{}{\Phi} \tilde{\psi}_{k}(x; \Phi)
  = \sum_{k'} \tilde{\psi}_{k'}(x; \Phi) \tilde{A}_{k',k}(\Phi)
  .
\end{equation}
The solution of this equation for an ``initial'' condition
$\tilde{\psi}_{k'}(x; \Phi')$, against a variation of $\Phi$ along a path
$C$ from $\Phi'$ to $\Phi''$, is 
\begin{equation}
  \label{eq:tildePsiByW}
  \tilde{\psi}_{k}(x; \Phi'')
  = \sum_{k'}\tilde{\psi}_{k'}(x; \Phi') W_{k',k}(C),
\end{equation}
where
\begin{equation}
  \label{eq:defW}
  W(C)
  \equiv
  \AntiTexp\left(-i \int_{C} \tilde{A}(\Phi) d\Phi\right)
  ,
\end{equation}
and $\AntiTexp$ is the anti-path-ordered exponential.

A change of the phase factors in eigenfunctions
\begin{equation}
  \label{eq:aBasisTransformation}
  \tilde\psi_k(x;\Phi) 
  \mapsto
  \tilde\psi_k(x;\Phi) e^{i\eta_k(\Phi)}
\end{equation}
induces the following changes~\cite{Cheon-EPL-85-20001,TANAKA-AP-85-1340}
\begin{align}
  \tilde{A}_{k''k'}(\Phi) 
  &
  \mapsto
  e^{-i[\eta_{k''}(\Phi)-\eta_{k'}(\Phi)]}\tilde{A}_{k''k'}(\Phi) 
  -\pdfrac{\eta_k(\Phi)}{\Phi}\delta_{k''k'}
  ,
  \\
  W_{k'',k'}(C)
  &
  \mapsto
  e^{-i\eta_{k''}(\Phi')}{W}_{k'',k'}(C)e^{i\eta_{k'}(\Phi'')}
  .
\end{align}
It is also straightforward to see the covariance of 
the holonomy matrix~\eqref{eq:holonomyMatrix}
against the above change 
\begin{align}
  M_{k'',k'}(C)
  &
  \mapsto 
  e^{-i\eta_{k''}(\Phi')}{M}_{k'',k'}(C)e^{i\eta_{k'}(\Phi')}
  .
\end{align}

An example of the evaluation of Eq.~\eqref{eq:defW} is shown.
Using a choice of the eigenfunction in Eq.~\eqref{eq:defTildePsik},
we obtain the gauge connection in the Byers-Yang gauge 
\begin{equation}
  \label{eq:APT}
  \tilde{A}_{k'',k'}(\Phi)
  = \frac{i}{k''-k'}(1-\delta_{k'',k'})
  ,
\end{equation}
which is denoted as $\tilde{A}_{k'',k'}$, since it happens to be
independent of $\Phi$.
We obtain the $W$-matrix~[Eq.~\eqref{eq:defW}] using a Fourier transformation
of $\tilde{A}_{k'',k'}$:
\begin{equation}
  \tilde{A}(\theta'',\theta')
  =\frac{1}{2\pi} \sum_{k''=-\infty}^{\infty}\sum_{k'=-\infty}^{\infty}
  e^{i k''\theta''} A_{k'',k'}e^{-i k'\theta'}
  .
\end{equation}
For $0\le \theta',\theta'' < 2\pi$, we have
\begin{equation}
  \tilde{A}(\theta'',\theta')
  = (\theta''-\pi)\delta(\theta''-\theta')
  ,
\end{equation}
from the Poisson summation formula~\cite{PSF},
\begin{equation}
  \sum_{k=-\infty}^{\infty} e^{i k\theta}
  = 2\pi\sum_{m=-\infty}^{\infty} \delta(\theta - 2\pi m)
  .
\end{equation}
Hence, we may say that 
$\tilde{A}$ is a diagonal operator in 
``$\theta$-representation''. Accordingly, we have
\begin{equation}
  [\exp(-ix\tilde{A})](\theta'',\theta')
  =
  e^{-i x (\theta''-\pi)}
  \delta(\theta''-\theta')
  .
\end{equation}
Using the inverse Fourier transformation, we obtain
\begin{align}
  \label{eq:resultW}
  W_{k'',k'}(C)
  =
  e^{-i(k''-k')\pi}
  \frac{\sin[\pi(\Phi''-\Phi'+k''-k')]}{\pi(\Phi''-\Phi'+k''-k')}
  .
\end{align}
For a periodic increment of $\Phi$, we obtain
\begin{equation}
  W_{k'',k'}(C) = e^{i\pi}\delta_{k''+1,k'}
  .
\end{equation}
This offers a way to obtain 
Eq.~(\ref{eq:parametricDependenceOnTildePsi}) 
from the non-Abelian gauge connection $\tilde{A}(\Phi)$.

We explain how the prescription with a non-Abelian connection is
inapplicable to the periodic gauge in Sec.~\ref{sec:singleValued}.
Because the eigenfunction $\psi_k(x)$ in the periodic gauge is
independent of $\Phi$ [see, Eq.~\eqref{eq:eigenfunctionInPeriodicGauge}],
the non-Abelian connection is trivial, i.e.,
\begin{align*}
  A_{k'',k'}(\Phi)
  &
  \equiv
  \langle {\psi}_{k''}(x), i\pdfrac{}{\Phi}{\psi}_{k'}(x)\rangle 
  \\ &
  = 0
  .
\end{align*}
Although $C$ is an open path under the periodic gauge, it is 
straightforward to extend the prescription to obtain the
holonomy matrix from the non-Abelian 
connection~\cite{Cheon-EPL-85-20001,TANAKA-AP-85-1340}.
Namely, for an adiabatic change of $\Phi$ along $C$, the holonomy matrix
is 
\begin{equation}
  \AntiTexp\left(-i \int_{C} {A}(\Phi) d\Phi\right)
  = 1
  ,
\end{equation}
where we use the fact that ${A}(\Phi)$ satisfies the parallel transport 
condition $A_{k,k}(\Phi) = 0$.
Hence the holonomy matrix in the periodic gauge has nothing to do
with the anholonomy, although this is consistent with the fact 
that $\psi_k(x)$ is independent of $\Phi$.

\section{A geometric significance of dynamical phase}
\label{sec:adiabaticBYgauge}
So far, we have examined the time evolution of
time-dependent Aharonov-Bohm ring in the Byers-Yang gauge rigorously, using 
the fact that eigenfunctions in the periodic gauge at each instant 
are independent of the magnetic flux. In this section,
we show an alternative analysis of the time evolution in 
the Byers-Yang gauge. We start from the time-dependent
Schr\"odinger equation in Byers-Yang gauge, and we keep
employing the Byers-Yang gauge throughout the evolution of the system.
A subtle point in the separation of geometric and dynamical phases is
to be elucidated.
On the other hand, due to the difficulty of the problem,
our analysis is restricted within the adiabatic time evolution.

We first consider the time-dependent Schr\"odinger equation 
in the Byers-Yang gauge.
In the following, we explicitly
denote the time-dependence of $A(x)$ as $A_t(x)$. 
Suppose $\tilde\psi(x,t)$ is obtained by the Byers-Yang gauge 
transformation [Eq.~\eqref{eq:BYgaugeTransformation}] of a time-dependent
wavefunction that satisfies the 
Schr\"odinger equation in the periodic gauge.
The time-dependent Schr\"odinger equation for $\tilde\psi(x,t)$
is 
\begin{equation}
  i\pdfrac{}{t}\tilde\psi(x,t)
  =
  \left[-\frac{1}{2}\pdfrac{^2}{x^2} + q\int_0^{x}\pdfrac{A_t(x')}{t} dx'\right]
  \tilde\psi(x,t)
  .
\end{equation}
Note that 
an extra term arises from the time-dependence of the vector potential.
Hence the time evolution of $\tilde\psi(x,t)$ is described by the Hamiltonian
\begin{equation}
  \label{eq:ExactHamiltonianInBYgauge}
  \tilde{H}(t)
  =
  -\frac{1}{2}\pdfrac{^2}{x^2}
  + \frac{{2\pi}x}{L}\frac{d\Phi(t)}{dt}
  ,
\end{equation}
where the second term represents the effect of 
the time-dependence of the magnetic flux~\cite{Avron-RMP-60-873}.

We now approximately solve the time-dependent Schr\"odinger equation
\begin{equation}
  i\pdfrac{}{t}\tilde{\Psi}_k(t)=\tilde{H}(t)\tilde{\Psi}_k(t)
\end{equation}
under the quasi-periodic boundary condition~\eqref{eq:BynersYangPBC}
and the initial condition $\tilde{\Psi}_k(t')=\tilde\psi_k(x; \Phi')$.
Here we assume that the magnetic flux adiabatically depends on time.
This justifies the assumption 
that $\tilde\psi_k(x; \Phi(t))$ approximates
well an eigenfunction of $\tilde{H}(t)$.
Because $\tilde\psi_k(x; \Phi)$ satisfies the parallel transport condition,
the final wavefunction, except its dynamical phase factor, is
$\tilde\psi_k(x; \Phi'')$~\cite{Simon-PRL-51-2167}.
In order to obtain a good approximation of the dynamical phase, 
on the other hand,
we need to take into account the leading correction of the
eigenenergy of $\tilde{H}(t)$
\begin{equation}
  \label{eq:tildeEk}
  \tilde{E}_k(t)
  = E_k[\Phi(t)] +\pi\frac{d\Phi(t)}{dt}
  ,
\end{equation}
where
\begin{equation}
  \tilde{E}_k(t) 
  \equiv
  \langle \tilde{\psi}_k(x, \Phi(t)),\;
  \tilde{H}(t)\tilde{\psi}_k(x, \Phi(t))\rangle
  .
\end{equation}
The second term of Eq.~\eqref{eq:tildeEk} offers a nontrivial correction to 
the dynamical phase.
Indeed, its integration 
\begin{equation}
  \label{eq:extraPhaseTerm}
  \int_{t'}^{t''}\pi\frac{d\Phi(t)}{dt} dt
  = \pi (\Phi'' - \Phi')
\end{equation}
does not vanish even in the adiabatic limit $d\Phi/dt\to0$.
This is the origin of the second factor in 
Eq.~\eqref{eq:exactTimeEvolutionViaPeriodicGauge}.
Although this factor is interpreted as a part of
dynamical phase factor in the Byers-Yang gauge,
this is classified as a part of the geometric factor
in the calculation through the periodic gauge.
In any case, we obtain Eq.~\eqref{eq:exactTimeEvolutionViaPeriodicGauge}
from the adiabatic time evolution in the Byers-Yang gauge.

Hence, the adiabatic time evolution of the quantum state 
whose initial
condition is the $k$-th eigenstate is 
\begin{align}
  \tilde{\Psi}_k(t'')
  &
  \equiv
  \exp\left[-i\int_{t'}^{t''}\tilde{E}_k(t) dt
    +i\int_{C} \tilde{A}_{kk}(\Phi)d\Phi\right]
  \nonumber\\ &\qquad{}\times
  \tilde{\psi}_k(x, \Phi'')
  ,
\end{align}
where $A_{kk}(\Phi)$ is Mead-Truhlar-Berry's gauge connection 
for the $k$-th eigenfunction 
$\tilde{\psi}_k(x, \Phi)$%
~\cite{Mead-JCP-70-2284,Berry-PRSLA-392-45,Samuel-PRL-60-2339}.
From Eq.~\eqref{eq:extraPhaseTerm}, we have
\begin{align}
  \tilde{\Psi}_k(t'')
  &
  =
  e^{i\gamma_{\rm D}-i\pi(\Phi''-\Phi')
  +i\int_{C} \tilde{A}_{kk}(\Phi)d\Phi}
  \tilde{\psi}_k(x, \Phi'')
  .
\end{align}
Excluding the dynamical phase in the periodic gauge $\gamma_{\rm D}$,
we define ``the geometric part'' of $\tilde{\Psi}_k(t'')$ as
\begin{align}
  \tilde{\Psi}^{({\rm g})}_k(t'')
  &
  =
  e^{-i\pi(\Phi''-\Phi')
  +i\int_{C} \tilde{A}_{kk}(\Phi)d\Phi}
  \tilde{\psi}_k(x, \Phi'')
  .
\end{align}
The holonomy matrix for the adiabatic approximation is 
\begin{align}
  M_{k'',k'}^{({\rm g})}
  \equiv
  \langle\tilde\psi_{k''}(x,\Phi'),
  \tilde{\Psi}^{({\rm g})}_{k'}(t'')\rangle
  .
\end{align}
In terms of the non-Abelian gauge connection $\tilde{A}(\Phi)$
[Eq.~\eqref{eq:gaugeConnectionInBYgauge}], we obtain
\begin{align}
  \label{eq:defMg}
  M_{k'',k'}^{({\rm g})}(C)
  &
  = e^{-i\pi(\Phi''-\Phi')}
  \left[\AntiTexp\left(-i \int_{C} \tilde{A}(\Phi) d\Phi\right)\right]_{k'',k'}
  \nonumber\\ &\qquad{}\times
  \exp\left[i\int_C \tilde{A}_{k'k'}(\Phi)d\Phi\right]
  ,
\end{align}
where Eqs.~\eqref{eq:tildePsiByW} and \eqref{eq:defW} are used.
In the conventional approach of the eigenspace anholonomy,
the holonomy matrix is described solely by the non-Abelian gauge 
connection $\tilde{A}(\Phi)$~\cite{Cheon-EPL-85-20001}.
However, $M_{k'',k'}^{({\rm g})}$ has an extra factor 
that comes from the dynamical phase in the Byers-Yang gauge.
This factor is required to keep the consistency with 
the analysis in Sec.~\ref{sec:BYgauge}. To see this,
we evaluate $M_{k'',k'}^{({\rm g})}$ using the choice
of eigenfunctions in Eq.~\eqref{eq:defTildePsik}.
From the evaluation of the $W$-matrix [Eq~\eqref{eq:resultW}] 
and the parallel transport condition $\tilde{A}_{k'k'}(\Phi)=0$
[see, Eq.~\eqref{eq:APT}], it is shown that
$M_{k'',k'}^{({\rm g})}(C)$ agrees with 
$M_{k'',k'}(C)$ [Eq.~\eqref{eq:holonomyMatrix}].

We finally remark that the eigenenergies of
the Hamiltonian [Eq.~\eqref{eq:ExactHamiltonianInBYgauge}] have
avoided crossings, which were ignored in the above arguments.
Because the spectral degeneracies in the unperturbed Hamiltonian are
lifted by the perturbation, i.e., the second term in 
Eq.~\eqref{eq:ExactHamiltonianInBYgauge}, we need to take into
account their effect on the adiabatic time evolution.
Here, the nonadiabatic transition across an avoided
crossing corresponds to the event that the system is kept to stay in
an approximate eigenstate $\tilde{\psi}_k(x, \Phi)$.
Because the magnitude of the perturbation is proportional to
$d\Phi/dt$, the nonadiabatic transition probability is 
unity in the adiabatic limit. This gives a justification to our
assumption of adiabatic change
along $\tilde{\psi}_k(x, \Phi(t))$, 
and as a result, also justifies our procedural assumption to neglect 
the seemingly possible occurrence of avoided crossing in the adiabatic limit.

\section{Summary and Discussion}
We have shown that the Aharonov-Bohm ring with a vanishing electrostatic 
potential offers an example of anholonomies of eigenvalue and eigenstates.
In particular, this system offers an extension of the quantum anholonomies
for the nonadiabatic regime.
At the same time, it is shown that the appearance of 
the eigenspace anholonomy depends on the choice of the vector 
potential of the magnetic flux.
It is also shown that the holonomy matrix, which is the central object
in the conventional prescription of the eigenspace anholonomy,
offers a sensible answer only under the Byers-Yang gauge. Although this 
is legitimate because all the anholonomy of the present example
is summarized as the eigenspace anholonomy under the Byers-Yang gauge, 
it is desirable to develop a prescription that is manifestly independent 
of the choice of the vector potential, instead of such an 
{\it ad hoc} argument. 
A possible strategy would be to develop the ``gauge theory'' for 
the anholonomies in 
eigenenergies, or in the collection of the expectation values of eigenstates. 
We leave this as an open question.
The conventional examples of eigenvalue and the eigenspace 
anholonomies requires a rank-$1$ perturbation and 
its cousins~\cite{Cheon-PLA-248-285,Tanaka-PRL-98-160407,%
Miyamoto-PRA-76-042115,Cheon-EPL-85-20001,Cheon-PLA-374-144}.
In contrast to this, the eigenspace anholonomy in the present example 
emerges from the quasi-periodic boundary condition. 
Thus another subtleness of the vector potential 
in the quantum theory~\cite{Aharonov-PR-115-485}
is revealed.

\section*{Acknowledgments}
AT wishes to thank Fumihiko Nakano for a useful conversation.
This work has been partially supported by the Grant-in-Aid for Scientific 
Research of MEXT, Japan (Grants No. 22540396 and 21540402).


%


\end{document}